\documentclass[aip,graphicx]{revtex4-1}

\draft % marks overfull lines with a black rule on the right

\begin{document}

\title{Second proof for $SU(2)$ isomorphic to $(\bigotimes LB1)^{3}$ theorem} %Title of paper

\author{Alcides Garat}
%\thanks{}
\affiliation{1. Instituto de F\'{\i}sica, Facultad de Ciencias, Igu\'a 4225, esq. Mataojo, Montevideo, Uruguay.}

\date{May 1st, 2012}

\begin{abstract}
In this note we present a second independent proof for the theorem introduced previously that establishes an isomorphism between $SU(2)$ and $(\bigotimes LB1)^{3}$. Since the local groups LB1 and LB2 are isomorphic, it was also previously proved a similar result for $(\bigotimes LB2)^{3}$. We are going to reverse the three sets of tetrads that are going to be used in order to prove this new version. Instead of choosing three $SU(2)$ different tetrads keeping the electromagnetic tetrad the same for all three sets of $SU(2)$ tetrads, we are going to keep fixed the $SU(2)$ tetrad, that is, we are going to pick just one $SU(2)$ tetrad but choose three different arbitrary but fixed electromagnetic tetrads in order to gauge locally the only local $SU(2)$ tetrad involved in our new version of this theorem. The same result will be obtained via an alternative but equivalent way.
\end{abstract}


\maketitle %\maketitle must follow title, authors, abstract and \pacs

\section{Introduction}

In this short note we are going to present an idea already studied in paper \cite{AYM} under a new light and in a different qualitative fashion. In manuscript \cite{AYM} it was proven that locally in a curved four-dimensional Lorentzian spacetime where electromagnetic and $SU(2)$ gauge fields are present, the local group $SU(2)$ is isomorphic to the tensor product of three local LB1 groups of transformations $(\bigotimes LB1)^{3}$. In order to prove this theorem we made use of three different sets of local $SU(2)$ tetrads whose gauge vectors involved just one electromagnetic local tetrad. In our new version this ideas are going to be reversed, we are going to use just a single local $SU(2)$ tetrad and three different local electromagnetic tetrads. To this end in section \ref{newgaugevectors} we are going to introduce three Lorentz rotated electromagnetic tetrads. That is, the Lorentz local transformations that relate the three sets of electromagnetic tetrads are local spatial transformations and these transformations are arbitrary but fixed. Through the concept of tetrad structure invariance we are going to present the three sets of tetrads built in the same fashion. That is, with similar skeleton and gauge vector structure. This elements are going to allow for the construction of the local gauge vector appropriate to our new version of the theorem. Finally we are going to study the new proof of this theorem and realize that the new gauge vector by its very constructive nature permits several ways of considering the same isomorphism.

\section{New gauge vectors}
\label{newgaugevectors}

We proceed then to pick up the only $SU(2)$ tetrad involved in our new proof. Following the same procedure depicted in manuscript \cite{AYM} section ``Extremal field in $SU(2)$ geometrodynamics'' we introduce using the same notation, the tetrad,

\begin{eqnarray}
S_{(1)}^{\mu} &=& \epsilon^{\mu\lambda}\:\epsilon_{\rho\lambda}\:X^{\rho}
\label{S1}\\
S_{(2)}^{\mu} &=& \sqrt{-Q_{ym}/2} \: \epsilon^{\mu\lambda} \: X_{\lambda}
\label{S2}\\
S_{(3)}^{\mu} &=& \sqrt{-Q_{ym}/2} \: \ast \epsilon^{\mu\lambda} \: Y_{\lambda}
\label{S3}\\
S_{(4)}^{\mu} &=& \ast \epsilon^{\mu\lambda}\: \ast\epsilon_{\rho\lambda}
\:Y^{\rho}\ ,\label{S4}
\end{eqnarray}

where $Q_{ym} = \epsilon_{\mu\nu}\:\epsilon^{\mu\nu}$ and $\epsilon_{\mu\nu}$ is new kind of local $SU(2)$ gauge invariant extremal tensor already introduced in paper \cite{AYM}. Once again we are going to call for future reference for instance $\epsilon^{\mu\lambda}\:\epsilon_{\rho\lambda}$ the skeleton of the tetrad vector $S_{(1)}^{\mu}$, and $X^{\rho}$ the gauge vector. In the case of $S_{(3)}^{\mu}$, the skeleton will be $\ast \epsilon^{\mu\lambda}$, and $Y_{\lambda}$ will be the gauge vector. Similar for the other tetrad vectors.  It is clear now that skeletons are gauge invariant by construction, under $SU(2) \times U(1)$, and again we cite paper \cite{AYM} for the proof. This property guarantees that the vectors under local $U(1)$ or $SU(2)$ gauge transformations are not going to leave their original planes or blades, keeping therefore the metric tensor explicitly invariant. Now the gauge vectors remain to be defined. They are the key to our new version of the theorem, but first we are going to introduce three new electromagnetic tetrads. The first electromagnetic tetrad we are going to introduce is the standard $E_{\alpha}^{\:\:\mu}$ introduced in paper \cite{A}. For the second electromagnetic tetrad we are going to need a local Lorentz transformation, specifically a local spatial rotation. In order to compare with the expressions in manuscript \cite{AYM} let us analyze the expression $\tilde{E}_{\delta}^{\:\:\rho} = \Lambda^{\alpha}_{\:\:\delta}\:E_{\alpha}^{\:\:\rho}$. This is going to be a Lorentz transformed electromagnetic tetrad vector. The Lorentz transformations are simple spatial rotations. Then, keeping the same notation as in \cite{A}, we call,

\begin{eqnarray}
\tilde{\xi}^{\mu\nu} &=& -2\:\sqrt{-Q/2}\:\Lambda^{\delta}_{\:\:o}\:\Lambda^{\gamma}_{\:\:1}\:E_{[\delta}^{\:\:\mu}\:E_{\gamma]}^{\:\:\nu}\label{ET}\\
\ast \tilde{\xi}^{\mu\nu} &=& 2\:\sqrt{-Q/2}\:\Lambda^{\delta}_{\:\:2}\:\Lambda^{\gamma}_{\:\:3}\:E_{[\delta}^{\:\:\mu}\:E_{\gamma]}^{\:\:\nu}\ .\label{DET}
\end{eqnarray}

Now, with these fields, the $\tilde{\xi}_{\mu\nu}$, and its dual $\ast \tilde{\xi}_{\mu\nu}$, we can repeat the procedure followed in \cite{A}, and the transformed tetrads $\tilde{E}_{\alpha}^{\:\:\rho}$, can be rewritten completely in terms of these ``new'' extremal fields. We repeat here for the sake of clarity a few remarks made in manuscript \cite{AYM}. It is straightforward to prove that $\tilde{\xi}^{\mu\lambda}\:\ast \tilde{\xi}_{\mu\nu} = 0$.
It is also evident that $\tilde{E}_{o}^{\:\:\mu} \:\ast \tilde{\xi}_{\mu\nu} = 0 = \tilde{E}_{1}^{\:\:\mu} \:\ast \tilde{\xi}_{\mu\nu}$. Therefore $\tilde{E}_{o}^{\:\:\mu}$ and $\tilde{E}_{1}^{\:\:\mu}$ belong to the plane generated by the normalized version of vectors like $\tilde{\xi}^{\mu\nu}\:\tilde{\xi}_{\lambda\nu}\:X^{\lambda}$ and $\tilde{\xi}^{\mu\nu}\:X_{\nu}$.
Then, for instance we are going to be able to write the timelike $\tilde{E}_{o}^{\:\:\mu}$ as the the normalized version of the timelike $\tilde{\xi}^{\mu\nu}\:\tilde{\xi}_{\lambda\nu}\:\tilde{X}^{\lambda}$ for some vector field  $\tilde{X}^{\lambda}$. We remind ourselves that the relation between the normalized versions of the two vectors that locally determine blade one, $\tilde{\xi}^{\mu\nu}\:\tilde{\xi}_{\lambda\nu}\:X^{\lambda}$ and  $\tilde{\xi}^{\mu\nu}\:X_{\nu}$ on one hand, and $\tilde{\xi}^{\mu\nu}\:\tilde{\xi}_{\lambda\nu}\:\tilde{X}^{\lambda}$ on the other hand, is established through a LB1 gauge transformation \cite{A}. Analogous analysis for $\tilde{E}_{2}^{\:\:\mu}$ and $\tilde{E}_{3}^{\:\:\mu}$ on blade two.  Gauge transformations of the electromagnetic tetrads we remind ourselves are nothing but a special kind of tetrad transformations that belong either to the groups LB1 or LB2. This method essentially says that the local Lorentz transformation of the electromagnetic tetrads is structure invariant, or construction invariant. This means that after a Lorentz transformation we can manage to rewrite the new transformed tetrads using skeletons and gauge vectors following the same pattern as for the original tetrad before the Lorentz transformation. We are going to call this property tetrad structure covariance. Having reviewed this method that allows for the construction of new electromagnetic tetrads which are all structure covariant, we introduce two more local electromagnetic tetrads $\overline{E}_{\delta}^{\:\:\rho}$ and $\hat{E}_{\delta}^{\:\:\rho}$ for two arbitrary but fixed spatial rotations. The rotations are different between them and not trivial of course. Tetrad $\overline{E}_{\delta}^{\:\:\rho}$ is defined by the new extremal field $\overline{\xi}_{\mu\nu}$ and gauge vectors $\overline{X}^{\mu} = A^{\mu}$ and $\overline{Y}^{\mu} = \ast A^{\mu}$. The other tetrad $\hat{E}_{\delta}^{\:\:\rho}$ is defined by the new extremal field $\hat{\xi}_{\mu\nu}$ and gauge vectors $\hat{X}^{\mu} = A^{\mu}$ and $\hat{Y}^{\mu} = \ast A^{\mu}$. Let us remember that in this particular case $\ast A^{\mu}$ is just a name and not a dual \cite{A}. Again we insist that the two sets of tetrads just introduced are the result of local spatial rotations with respect to the standard electromagnetic tetrad, rewritten exactly as the original electromagnetic tetrad due to structure covariance for new local extremal fields, using the same local gauge vectors as the original tetrad. Now, we already knew that the structure $E_{\alpha}^{\:\:[\rho}\:E_{\beta}^{\:\:\lambda]}\:\ast \xi_{\rho\sigma}\:\ast \xi_{\lambda\tau}$ is invariant under $U(1)$ local gauge transformations. Essentially, because of the electromagnetic extremal field property \cite{A}$^{,}$\cite{MW}, $\xi_{\mu\sigma}\:\ast \xi^{\mu\tau} = 0$. Likewise and for the same reason, the structures $\overline{E}_{\alpha}^{\:\:[\rho}\:\overline{E}_{\beta}^{\:\:\lambda]}\:\ast \overline{\xi}_{\rho\sigma}\:\ast \overline{\xi}_{\lambda\tau}$ and $\hat{E}_{\alpha}^{\:\:[\rho}\:\hat{E}_{\beta}^{\:\:\lambda]}\:\ast \hat{\xi}_{\rho\sigma}\:\ast \hat{\xi}_{\lambda\tau}$ are invariant under $U(1)$ local gauge transformations as well. With all these elements we can proceed now to the construction of the new local gauge vectors that are going to be relevant to our new version of the theorem. We are going to use the three new defined electromagnetic tetrads in the construction of the new gauge vectors in such a way that when we perform a local $SU(2)$ gauge transformation, three commuting and independent LB1 Abelian transformations are going to be generated on the same blade either one or two. Not on three different planes, either one or two associated to three different tetrad sets as the first proof in manuscript \cite{AYM}. Let us remind ourselves that the tetrad that we defined in (\ref{S1}-\ref{S4}) was just one $SU(2)$ tetrad with just one associated blade one and just one blade two. There lies the essence of this new version. We proceed then to define,

\begin{eqnarray}
X^{\sigma} = Tr[\Sigma^{\alpha\beta}\: ( E_{\alpha}^{\:\:\rho}\: E_{\beta}^{\:\:\lambda}\:\ast \xi_{\rho}^{\:\:\sigma}\:\ast \xi_{\lambda\tau} +
\overline{E}_{\alpha}^{\:\:\rho}\: \overline{E}_{\beta}^{\:\:\lambda}\:\ast \overline{\xi}_{\rho}^{\:\:\sigma}\:\ast \overline{\xi}_{\lambda\tau} +
\hat{E}_{\alpha}^{\:\:\rho}\: \hat{E}_{\beta}^{\:\:\lambda}\:\ast \hat{\xi}_{\rho}^{\:\:\sigma}\:\ast \hat{\xi}_{\lambda\tau} ) \:A^{\tau}]\ . \label{NEWGAUGEVECTORS}
\end{eqnarray}

We also define $X^{\sigma} = Y^{\sigma}$. The object $\Sigma^{\alpha\beta}$ was explained in detail in Appendix II of manuscript \cite{AYM}. It translates local $SU(2)$ gauge transformations in $A^{\tau}$ into local spatial Lorentz rotations.

%\nonumber \\

\section{The theorem new version}
\label{theonewversion}

Let us study now the transformation properties of the object (\ref{NEWGAUGEVECTORS}) under the local $SU(2)$ gauge transformation $S$,

\begin{eqnarray}
A_{\mu} \: \rightarrow S^{-1}\:A_{\mu}\:S + {\imath \over g} S^{-1}\:\partial_{\mu}(S) \ . \label{GTA}
\end{eqnarray}

Therefore, the gauge vectors (\ref{NEWGAUGEVECTORS}) transform as,

\begin{eqnarray}
Tr[\Sigma^{\alpha\beta}\:( E_{\alpha}^{\:\:\rho}\: E_{\beta}^{\:\:\lambda}\:\ast \xi_{\rho}^{\:\:\sigma}\:\ast \xi_{\lambda\tau} +
\overline{E}_{\alpha}^{\:\:\rho}\: \overline{E}_{\beta}^{\:\:\lambda}\:\ast \overline{\xi}_{\rho}^{\:\:\sigma}\:\ast \overline{\xi}_{\lambda\tau} +
\hat{E}_{\alpha}^{\:\:\rho}\: \hat{E}_{\beta}^{\:\:\lambda}\:\ast \hat{\xi}_{\rho}^{\:\:\sigma}\:\ast \hat{\xi}_{\lambda\tau} )\:A^{\tau}] \: \rightarrow \nonumber \\ Tr[\Sigma^{\alpha\beta}\:( E_{\alpha}^{\:\:\rho}\: E_{\beta}^{\:\:\lambda}\:\ast \xi_{\rho}^{\:\:\sigma}\:\ast \xi_{\lambda\tau} +
\overline{E}_{\alpha}^{\:\:\rho}\: \overline{E}_{\beta}^{\:\:\lambda}\:\ast \overline{\xi}_{\rho}^{\:\:\sigma}\:\ast \overline{\xi}_{\lambda\tau} +
\hat{E}_{\alpha}^{\:\:\rho}\: \hat{E}_{\beta}^{\:\:\lambda}\:\ast \hat{\xi}_{\rho}^{\:\:\sigma}\:\ast \hat{\xi}_{\lambda\tau} )\:S^{-1}\:A^{\tau}\:S] +
 \nonumber \\
{\imath \over g}\: Tr[\Sigma^{\alpha\beta}\:( E_{\alpha}^{\:\:\rho}\: E_{\beta}^{\:\:\lambda}\:\ast \xi_{\rho}^{\:\:\sigma}\:\ast \xi_{\lambda\tau} +
\overline{E}_{\alpha}^{\:\:\rho}\: \overline{E}_{\beta}^{\:\:\lambda}\:\ast \overline{\xi}_{\rho}^{\:\:\sigma}\:\ast \overline{\xi}_{\lambda\tau} +
\hat{E}_{\alpha}^{\:\:\rho}\: \hat{E}_{\beta}^{\:\:\lambda}\:\ast \hat{\xi}_{\rho}^{\:\:\sigma}\:\ast \hat{\xi}_{\lambda\tau} )\:S^{-1}\:\partial^{\tau}\:(S)] \ \label{XCHOICE}
\end{eqnarray}

The field strength transforms as usual, $f_{\mu\nu} \rightarrow S^{-1}\:f_{\mu\nu}\:S$, and similar for the extremal $\xi_{\mu\nu}$, and their duals. Then, we can follow exactly the same guidelines laid out in \cite{A}, to study the gauge transformations of the tetrad vectors on blades one, and two. We can carry over into the present work, all the analysis done in the gauge geometry section in \cite{A}. It is clear that the vectors $S_{(1)}^{\mu}$ and $S_{(2)}^{\mu}$, by virtue of their own construction, remain on blade one after the (\ref{GTA}) transformation. It is also evident that after the transformation they are orthogonal. These two facts mean that the metric tensor is invariant under the transformations (\ref{GTA}) when the two vectors are normalized. We are assumming for simplicity that $S_{(1)}^{\mu}$ is timelike and $S_{(2)}^{\mu}$ spacelike, both vectors non-trivial. After this point we can repeat exactly all the steps in section ``Gauge geometry'' of manuscript \cite{AYM}. We can study the transformation properties of the two vectors $S_{(1)}^{\mu}$ and $S_{(2)}^{\mu}$ on blade one and similar for $S_{(3)}^{\mu}$ and $S_{(4)}^{\mu}$ on blade two. Then we can repeat all the analysis of the memory of the transformation, etc. We are interested now in the result of this local gauge transformations for the vectors $S_{(1)}^{\mu}$ and $S_{(2)}^{\mu}$ with regard to the theorem we want to prove. For the vectors $S_{(3)}^{\mu}$ and $S_{(4)}^{\mu}$ spanning blade two the analysis would be analogous. Therefore we skip all the steps already studied in paper \cite{AYM} and go directly to the following expression for the locally transformed gauge vector,

\begin{eqnarray}
X^{\sigma} &\rightarrow& Tr[\tilde{\Lambda}^{\alpha}_{\:\:\:\delta}\:\tilde{\Lambda}^{\beta}_{\:\:\:\gamma}\:\Sigma^{\delta\gamma}\:
(E_{\alpha}^{\:\:\rho}\: E_{\beta}^{\:\:\lambda}\:\ast \xi_{\rho}^{\:\:\sigma}\:\ast \xi_{\lambda\tau} +
\overline{E}_{\alpha}^{\:\:\rho}\: \overline{E}_{\beta}^{\:\:\lambda}\:\ast \overline{\xi}_{\rho}^{\:\:\sigma}\:\ast \overline{\xi}_{\lambda\tau}  + \hat{E}_{\alpha}^{\:\:\rho}\: \hat{E}_{\beta}^{\:\:\lambda}\:\ast \hat{\xi}_{\rho}^{\:\:\sigma}\:\ast \hat{\xi}_{\lambda\tau} )\:    \:A^{\tau}]  +  \nonumber \\ &&{\imath \over g} \: Tr[\tilde{\Lambda}^{\alpha}_{\:\:\:\delta}\:\tilde{\Lambda}^{\beta}_{\:\:\:\gamma}\:\Sigma^{\delta\gamma}\:
E_{\alpha}^{\:\:\rho}\: E_{\beta}^{\:\:\lambda}\:\ast \xi_{\rho}^{\:\:\sigma}\:\ast \xi_{\lambda\tau}\:\partial^{\tau}\:(S)\:S^{-1}] +  \nonumber \\ &&{\imath \over g} \: Tr[\tilde{\Lambda}^{\alpha}_{\:\:\:\delta}\:\tilde{\Lambda}^{\beta}_{\:\:\:\gamma}\:\Sigma^{\delta\gamma}\:
\overline{E}_{\alpha}^{\:\:\rho}\: \overline{E}_{\beta}^{\:\:\lambda}\:\ast \overline{\xi}_{\rho}^{\:\:\sigma}\:\ast \overline{\xi}_{\lambda\tau} \:\partial^{\tau}\:(S)\:S^{-1}] +  \nonumber \\ &&{\imath \over g} \: Tr[\tilde{\Lambda}^{\alpha}_{\:\:\:\delta}\:\tilde{\Lambda}^{\beta}_{\:\:\:\gamma}\:\Sigma^{\delta\gamma}\:
\hat{E}_{\alpha}^{\:\:\rho}\: \hat{E}_{\beta}^{\:\:\lambda}\:\ast \hat{\xi}_{\rho}^{\:\:\sigma}\:\ast \hat{\xi}_{\lambda\tau} \:\partial^{\tau}\:(S)\:S^{-1}]
\label{XT}
\end{eqnarray}

For the sake of simplicity we are using the notation, $\Lambda^{(-1)\,\alpha}_{\:\:\:\:\:\:\:\:\:\:\:\:\delta} =\tilde{\Lambda}^{\alpha}_{\:\:\:\delta}$ and no confusion should arise for instance with $\tilde{E}_{\delta}^{\:\:\rho} = \Lambda^{\alpha}_{\:\:\delta}\:E_{\alpha}^{\:\:\rho}$. In expression (\ref{XT}) and building on our experience with previous similar formulas in manuscripts \cite{A} and \cite{AYM} we know that the second line is associated to a LB1 local transformation on blade one that we are going to characterize by an angle $\phi$, the third line to a new LB1 transformation and independent from the previous transformation $\phi$ that we are going to characterize by the angle $\overline{\phi}$ and the forth line to a new LB1 transformation independent from $\phi$ and $\overline{\phi}$ that we are going to associate to the angle $\hat{\phi}$. All these three LB1 transformations are taking place locally on blade one and commute among themselves. The key to their independence are the rotated electromagnetic tetrads $\overline{E}_{\beta}^{\:\:\lambda}$ and $\hat{E}_{\beta}^{\:\:\lambda}$ with respect to the standard $ E_{\beta}^{\:\:\lambda}$ inside the gauge vector. The rotations are arbitrary but fixed, therefore, the three generated LB1 transformations associated to the angles $\phi$, $\overline{\phi}$ and $\widehat{\phi}$ are independent among themselves. It was already proven in manuscript \cite{AYM} in the section ``Gauge geometry'' that the mapping associated to local gauge transformations of the new tetrads generated by (\ref{GTA}) is surjective, injective, the image of this group mapping is not a subgroup of the LB1 group and that these transformations have no memory, therefore we are able now to formulate the following results,

\newtheorem {guesslb1} {Theorem}
\newtheorem {guesslb2}[guesslb1] {Theorem}
\begin{guesslb1}
The mapping between the local gauge group $SU(2)$ of transformations and the tensor product of the three local groups of LB1 tetrad transformations is isomorphic. \end{guesslb1}

Following analogously the reasoning laid out in \cite{A}, in addition to the ideas above, we can also state,

\begin{guesslb2}
The mapping between the local gauge group $SU(2)$ of transformations and the tensor product of the three local groups of LB2 tetrad transformations is isomorphic. \end{guesslb2}

\section{Conclusions}

There are several ways to visualize the isomorphism just studied. Let us write for instance expression (\ref{XT}) in a different order,

\begin{eqnarray}
X^{\sigma} &\rightarrow& Tr[\tilde{\Lambda}^{\alpha}_{\:\:\:\delta}\:\tilde{\Lambda}^{\beta}_{\:\:\:\gamma}\:\Sigma^{\delta\gamma}\:
E_{\alpha}^{\:\:\rho}\: E_{\beta}^{\:\:\lambda}\:\ast \xi_{\rho}^{\:\:\sigma}\:\ast \xi_{\lambda\tau}\:A^{\tau}]  +  \nonumber \\ &&
{\imath \over g} \: Tr[\tilde{\Lambda}^{\alpha}_{\:\:\:\delta}\:\tilde{\Lambda}^{\beta}_{\:\:\:\gamma}\:\Sigma^{\delta\gamma}\:
E_{\alpha}^{\:\:\rho}\: E_{\beta}^{\:\:\lambda}\:\ast \xi_{\rho}^{\:\:\sigma}\:\ast \xi_{\lambda\tau}\:\partial^{\tau}\:(S)\:S^{-1}] +  \nonumber \\ &&
Tr[\tilde{\Lambda}^{\alpha}_{\:\:\:\delta}\:\tilde{\Lambda}^{\beta}_{\:\:\:\gamma}\:\Sigma^{\delta\gamma}\:
\{\overline{E}_{\alpha}^{\:\:\rho}\: \overline{E}_{\beta}^{\:\:\lambda}\:\ast \overline{\xi}_{\rho}^{\:\:\sigma}\:\ast \overline{\xi}_{\lambda\tau}\:A^{\tau} +
{\imath \over g} \: 
\overline{E}_{\alpha}^{\:\:\rho}\: \overline{E}_{\beta}^{\:\:\lambda}\:\ast \overline{\xi}_{\rho}^{\:\:\sigma}\:\ast \overline{\xi}_{\lambda\tau} \:\partial^{\tau}\:(S)\:S^{-1}\}] +  \nonumber \\ &&
Tr[\tilde{\Lambda}^{\alpha}_{\:\:\:\delta}\:\tilde{\Lambda}^{\beta}_{\:\:\:\gamma}\:\Sigma^{\delta\gamma}\:
\{\hat{E}_{\alpha}^{\:\:\rho}\: \hat{E}_{\beta}^{\:\:\lambda}\:\ast \hat{\xi}_{\rho}^{\:\:\sigma}\:\ast \hat{\xi}_{\lambda\tau}\:A^{\tau} +
{\imath \over g} \: 
\hat{E}_{\alpha}^{\:\:\rho}\: \hat{E}_{\beta}^{\:\:\lambda}\:\ast \hat{\xi}_{\rho}^{\:\:\sigma}\:\ast \hat{\xi}_{\lambda\tau} \:\partial^{\tau}\:(S)\:S^{-1}\}]
\label{XTALT}
\end{eqnarray}

In expression (\ref{XTALT}) the second line represents a local LB1 transformation that we might call $\psi$, the third line a local LB1 transformation that we can call $\overline{\psi}$ and the forth line a local LB1 transformation that we would like to call $\widehat{\psi}$. This is a different way of assigning LB1 transformations. The theorem would provide the same conclusions. We can offer different sets of local LB1 transformations with associated local scalar angles like $\phi, \overline{\phi}, \widehat{\phi}$ or $\psi, \overline{\psi}, \widehat{\psi}$ and the reason resides in the commutativity of all these transformations whose Abelianity we can mathematically pinpoint in the additivity of gauge vectors. The physical conclusions associated to this theorem have been stated in detail in manuscript \cite{AYM}. We conclude by quoting from \cite{KK}, ``There is another stream of ideas flowing across the territory of the general theory of relativity. We can summarize its basic tenet by saying that all existing things are made out of geometry. Platos's vision that the world consists of four perfect solids, Clifford's picture of particles as wave packets of geometry, Einstein's whole life quest for a unified field theory, and Wheeler's conception of elementary particles as geometrodynamical excitons, follow this stream down from antiquity to the present time''.

%\bibliography{your-bib-file} % place the references here.

\end{document}